\documentclass[aps,prb,amsmath,showpacs,amssymb,amsfonts,10pt]{revtex4-1}

\usepackage{graphicx}
\usepackage{bm}

\newcommand{\half}{\dfrac{1}{2}}

\newcommand{\beq}{\begin{eqnarray}}
\newcommand{\eeq}{\end{eqnarray}}

\newcommand{\bfig}{ \left. \begin{array}{l} }
\newcommand{\bfigor}{ \left\{ \begin{array}{ll} }
\newcommand{\efig}{ \end{array} \right. }

\newcommand{\cH}{{\cal H}}
\newcommand{\cD}{{\cal D}}
\newcommand{\bS}{{\bf S}}

\newcommand{\non}{\nonumber}

\newcommand{\w}{\omega}
\newcommand{\e}{\epsilon}

\newcommand{\Sig}{\Sigma}

\newcommand{\Ga}{\Gamma}

\newcommand{\krest}{\dagger}
\newcommand{\akrest}{a^\krest}
\newcommand{\bkrest}{b^\krest}
\newcommand{\akr}{\akrest}
\newcommand{\bkr}{\bkrest}
\newcommand{\kk}{{\bf k}}
\newcommand{\qq}{{\bf q}}
\newcommand{\pp}{{\bf p}}
\newcommand{\kkk}{{{\kk}_0}}
\newcommand{\ppp}{{{\pp}_0}}

\begin{document}

\title{Quantum magnets with large single-ion easy-plane anisotropy in transverse magnetic field} 
\author{A.V. Sizanov$^1$}
\email{alexey.sizanov@gmail.com}
\author{A.V. Syromyatnikov$^{1,2}$}
\email{syromyat@thd.pnpi.spb.ru}
\affiliation{$^1$Petersburg Nuclear Physics Institute, Gatchina, St.\ Petersburg 188300, Russia}
\affiliation{$^2$Department of Physics, St.\ Petersburg State University, 198504 St.\ Petersburg, Russia}

\date{\today}

\begin{abstract}

We discuss low-temperature properties of magnets with integer spin and large single-ion easy-plane anisotropy $D$ in transverse magnetic field $h$. Considering the exchange interaction between spins as a perturbation and using the diagram technique we derive at $h\sim D$ in the first nonvanishing orders of the perturbation theory thermal corrections to the elementary excitation spectrum, magnetization and specific heat. An expression for the boundary $h_{c1}(T)$ is found in the $h$-$T$ plane between the paramagnetic phase and that with the long range magnetic order. The effective interaction is derived between bosons near the quantum critical point $h_{c1}(0)$. The proposed theory describes well experimental data obtained in $\rm NiCl_2$-$\rm 4SC(NH_2)_2$ (DTN).

\end{abstract}

\pacs{75.10.Jm, 75.40.Gb}

\maketitle

\underline{\it Introduction.} 
The topic of quantum criticality has received much attention in recent two decades. Of particular interest are quantum critical points (QCPs) which can be reached in experiments by varying easily controllable parameters such as external magnetic field, pressure, level of doping, etc. The equivalence between a spin system and a diluted gas of bosonic particles proved to be very useful in describing field-induced QCPs in magnets. \cite{giam} This equivalence is revealed and exploited using appropriate representation of spin operators via bosonic ones. 

We discuss in the present paper properties of a system on a 3D lattice with an integer spin and large single-ion anisotropy which is described by the Hamiltonian
\beq
\label{ham}
\cH= D \sum_{i} (S^{z}_{i})^{2}  + \frac12 \sum_{i,j} J_{i,j} \bS_{i} \bS_{j} + h \sum_{i} S^{z}_{i},
\eeq
where $D>0$ is assumed to be much larger than exchange constants ($D\gg J$) so that the ground state at $h=0$ is paramagnetic (all spins are approximately in the state with $S^z=0$). This system has at least two field-induced QCPs corresponding to transitions from the paramagnetic (at $h=h_{c1}(T=0)$) and from the fully polarized (at $h=h_{c2}(T=0)$) phases to other phases which nature depends on the detail of the exchange coupling and the lattice geometry. 
We propose in our recent paper \cite{larged1} a bosonic integer spin representation that is convenient for discussion of the system \eqref{ham} properties in the paramagnetic phase. Using this representation and considering the exchange interaction as a perturbation we find in Ref.~\cite{larged1} the spectrum of the Hamiltonian \eqref{ham} in the paramagnetic phase at $h=0$ in the third order in the perturbation theory (hereafter referred to as expansion in terms of $J/D$ for shot). We continue our study of the model \eqref{ham} in the present paper and address its low-temperature properties in the vicinity of the QCP $h=h_{c1}(T=0)$ using the proposed bosonic spin representation. Expressions are derived below in the first nonvanishing orders in $J/D$ for thermal corrections to the elementary excitation spectrum, magnetization and specific heat. An expression is found for $h_{c1}(T)$ that is the boundary of the paramagnetic phase in the $h$-$T$ plane. The effective interaction is derived between bosons near the QCP which can be observed experimentally. We demonstrate that the proposed theory describes well corresponding experimental data obtained in $\rm NiCl_2$--$\rm 4SC(NH_2)_2$ (DTN) \cite{add16,add17,12,13,14,yin,17,chern,add7,add8,add10,carlin,add14,15,sound,htrans} which is the most extensively studied compound having the paramagnetic ground state at $h=0$ and which is modeled by the Hamiltonian \eqref{ham}. 

The magnetic subsystem of DTN consists of Ni ions with $S=1$ and the Lande factor $g=2.26$. Magnetic ions form a body-centered tetragonal lattice which can be viewed as two interpenetrating tetragonal sublattices. The exchange interaction between spins inside one sublattice is antiferromagnetic and strongly anisotropic: the exchange constant along the tetragonal hard axis ($z$ axis) is much larger than those along $x$ and $y$ axes. Then, DTN is a quasi-1D material having two QCPs at 
\footnote{
\label{foot}
It should be noted that there is a certain discrepancy in values of $h_{c2}(0)$ in the experimental literature on DTN. Specific heat and magnetocaloric effect measurements give $h_{c2}(0)\approx12.6$~T. \cite{chern,13} On the other hand magnetization measurements \cite{add8} gives the value of $h_{c2}(0)$ very close to Eq.~\eqref{hc2dtn} obtained in ac susceptibility measurements \cite{yin}; anomalies in the sound velocity \cite{sound}, the sound attenuation \cite{sound} and the thermal conductivity \cite{chern,htrans} at a given $T<0.5$~K were observed at fields smaller than those in the specific heat\cite{chern}. The origin of this discrepancy is not discussed in the  literature and we have no explanation for this situation either. We choose in our recent \cite{larged1} and the present consideration of DTN the value \eqref{hc2dtn} because it provides better agreement between our expressions for the spectrum found in the third order in $J/D$ and the experimentally observed spectrum \cite{13} (see Ref.~\cite{larged1} for detail).}
\begin{eqnarray}
\label{hc1dtn}
 g\mu_Bh_{c1}^{DTN}(T=0) &=& 2.05\,{\rm T}, \\
\label{hc2dtn}
 g\mu_Bh_{c2}^{DTN}(T=0) &=& 12.175\, {\rm T}
\end{eqnarray}
with a canted antiferromagnetic phase between them. It was found \cite{13,yin} that the QCP $h= h_{c1}(0)$ belongs to the 3D BEC universality class: $h_{c1}(T)-h_{c1}(0)\propto T^\alpha$ with $\alpha\approx1.5$. The strength of the effective interaction between long-wavelength boson was extracted in Ref.\cite{17} from the measurements of magnetization and $h_{c1}(T)$. The specific heat at small $T$ was measured in Refs.~\cite{chern} at $h\sim h_{c1}(0)$ and $h\sim h_{c2}(0)$.

We have shown in our previous paper \cite{larged1} that Hamiltonian \eqref{ham} describes well the experimentally obtained spectrum of elementary excitations in DTN at $h=0$ with the following set of parameters which differs from the conventional one (see discussion in Ref.~\cite{larged1}):
\begin{eqnarray} 
\label{ourdtnpar}
 D &=& 7.72 \; \rm{K}, \non \\
 J_z &=& 1.86\; {\rm K}, \\
 J_{xy} &=& 0.2\; {\rm K}, \non\\
 V &=& 0.1\; \rm{K}, \non
\end{eqnarray}
where $J_{z}$ is the exchange constant along the chains, $J_{xy}$ is the exchange coupling constants between chains inside one tetragonal sublattice and $V$ is the exchange coupling constant between neighboring spins from different tetragonal sublattices (which is proposed in Ref.~\cite{prev}). We use these parameters below for the experimental data analysis.

\underline{\it Method and technique.} It is convenient to use the following spin representation of integer $S$ which is proposed in our previous study: \cite{larged1}
\begin{eqnarray}
 \label{sz} 
 S^z_i &=& \bkr_i b_i - \akr_i a_i, \\
\label{s+} 
S^+_i &=& S^x_i + i S^y_i = \bkr_i \sqrt{\dfrac{ ( S - \bkr_i b_i ) ( S+1 + \bkr_i b_i) }{ 1 + \bkr_i b_i }} +
		\sqrt{\dfrac{ ( S - \akr_i a_i ) ( S+1 + \akr_i a_i) }{ 1 + \akr_i a_i }} \cdot a_i,  \non\\
		&\approx& \bkr_i \left( c_1 - c_2 \; \bkr_i b_i \right) + \left( c_1 - c_2 \; \akr_i a_i \right) a_i
\end{eqnarray}
where $a_i$ and $b_i$ are bosonic operators, $c_1 = \sqrt{ S(S+1) }$ and $c_2  = \sqrt{ S(S+1) } - \sqrt{ (S-1)(S+2)/2 } > 0$.
Representation \eqref{sz}--\eqref{s+} reproduces the spin commutation relations on the physical subspace which is constrained by the following additional term in the Hamiltonian (see discussion in Ref.~\cite{larged1}):
\begin{equation}
\label{u} 
{\cal H}_{U} = \frac UN \sum_i  \akr_i \bkr_i a_i b_i, \quad U \to + \infty. 
\end{equation}
Substituting Eqs.~\eqref{sz}--\eqref{s+} into Eq.~\eqref{ham} and taking into account Eq.~\eqref{u} one obtains for the Hamiltonian
\begin{subequations}
\label{ham1}
\beq
\cH &=& 
\sum_{\pp} \left[ \e_{1a}(\pp) \akr_{\pp} a_{\pp} + \e_{1b}(\pp) \bkr_{\pp} b_{\pp} \right]
+ 
\sum_\pp \frac{ c_1^2 }{ 2 } J_\pp \left( \akr_\pp \bkr_{-\pp} + a_\pp b_{-\pp} \right) \\
\label{h4a}
&&{} + \frac1N \sum_{\pp_1+\pp_2=\pp_3+\pp_4} \left\{ \left[ D + \half J_{3-1} - \frac{ c_1 c_2 }{ 2 } \left( J_1 + J_3 \right) \right] \left( \akr_1 \akr_2 a_3 a_4 + \bkr_1 \bkr_2 b_3 b_4 \right) + \left[ U - J_{3-1} \right] \akr_1 \bkr_2 a_3 b_4  \right\} \\
\label{abbb}
&&{} - \frac1N \sum_{\pp_1+\pp_2+\pp_3=\pp_4} \frac{ c_1 c_2 }{ 2 } J_1 \left( \bkr_1 \akr_2 \akr_3 a_4 + \akr_1 \bkr_2 \bkr_3 b_4 + \akr_4 a_3 a_2 b_1 + \bkr_4 b_3 b_2 a_1 \right),
\eeq
\end{subequations}
where $ J_\pp = \sum_j J_{ij} e^{i \pp {\bf R}_{ij}} $, $N$ is the number of unit cells,
\beq
\label{e1a}
	\e_{1a,h}(\pp) &=& D + \frac{ c_1^2 }{ 2 } J_\pp-h,\\
\label{e1b}
	\e_{1b,h}(\pp) &=& D + \frac{ c_1^2 }{ 2 } J_\pp+h
\eeq
are spectra of $a$ and $b$ particles in the first order in $J/D$ (here and below the number in the lower index of the spectrum indicates its order in $J/D$). It is convenient for the following to introduce Green's functions
\begin{subequations}
\label{gf}
\beq
\label{gfa}
G_{a,h}(p) &=& -i \langle a_p \akr_p \rangle,\\
G_{b,h}(p) &=& -i \langle b_p \bkr_p \rangle,\\
F_h(p) &=& -i \langle \bkr_{-p} \akr_p \rangle,
\eeq
\end{subequations}
where $p=(\w,\pp)$ and $a_p$ is the Fourier transform of $a_\pp (\tau)$. Dyson equations for one couple of these Green's functions have the form
\begin{eqnarray}
\label{dyson}
 G_{a,h} ( p ) &=& G_{0a} ( p ) \left[ 1 + \Sig_{a,h}(p) G_{a,h} ( p ) + \Pi_h(p) F_h ( p ) \right], \\
 F_h ( p ) &=& G_{0b} ( -p) \left[ \overline{\Pi}_h(p) G_{a,h} ( p ) + \Sig_{b,h}(-p) F_h ( p ) \right],\non
\end{eqnarray}
where $ G_{0a} ( p ) = ( \w - \e_{1a,h}(\pp)+ i\delta )^{-1} $, $ G_{0b} ( p ) = ( \w - \e_{1b,h}(\pp)+ i\delta )^{-1} $, $\Sigma$ and $\Pi$ are normal and anomalous self-energy parts, respectively. Solving Eqs.~\eqref{dyson} and the couple of equations for $G_{b,h} ( p )$ and $F_h ( p )$ one obtains
\beq
\label{ga}
G_{a,h}(p) &=& \frac{ \w + \Sig_{b,h}(-p)}{\cD(p)},\\
\label{gb}
G_{b,h}(p) &=& \frac{ \w + \Sig_{a,h}(-p)}{\cD(-p)},\\
\label{f}
F_h (p)  &=& -\frac{ \overline{\Pi}_h(p) }{\cD(p)},\\
\label{den}
\cD (p) &=& \left(\w - \e_{1a}(\pp) - \Sig_{a,h} (p) \right)\left( \w + \e_{1b}(\pp) + \Sig_{b,h} (-p) \right) + |\Pi_h(p)|^2. 
\eeq
Spectra of $a$ and $b$ particles are given by equations
\beq
\label{dispeq}
\cD(\e_{a,h}(\pp),\pp) = 0,\qquad
\cD(-\e_{b,h}(\pp),\pp) = 0.
\eeq

\underline{\it $T=0$ and $h=0$.} Because $a$ and $b$ particles are equivalent at $h=0$ we have 
\begin{subequations}
\label{h=0}
\beq
\label{gh=0}
G_{a,h=0}(p)&=&G_{b,h=0}(p)=G(p),\\
\label{sigh=0}
\Sigma_{a,h=0}(p)&=&\Sigma_{b,h=0}(p)=\Sigma(p),\\
\Pi_{h=0}(p)&=&\Pi(p),\\
\label{eh=0}
\e_{a,h=0}(\pp)&=&\e_{b,h=0}(\pp)=\e(\pp).
\eeq
\end{subequations}
We calculate in our previous paper \cite{larged1} $\Sigma(p)$, $\Pi(\pp)$ and the spectrum $\e(\pp)$ up to the third order in $J/D$.

\underline{\it $T=0$ and $h\ne0$.} 
Taking into account that $[ \cH, \sum_iS_i^z ] =0$ and using Eq.~\eqref{sz} one concludes that $h$ and $-h$ play the role of chemical potentials for $a$ and $b$ particles, respectively, so that 
\begin{subequations}
\label{hne0}
\beq
\label{gahne0}
G_{a,h} ( \w,\pp ) &=& G ( \w + h,\pp  ), \\ 
\label{gbhne0}
G_{b,h} ( \w,\pp  ) &=& G ( \w - h,\pp  ), \\
F_{h} ( \w,\pp  ) &=& F (\w + h,\pp  ), 
\eeq
\end{subequations}
and, correspondingly, 
$\Sig_{a,h} ( \w,\pp  ) = \Sig (\w + h,\pp )$, 
$\Sig_{b,h} ( \w,\pp  ) = \Sig (\w - h,\pp )$,
$\Pi_{h} ( \w,\pp  ) = \Pi ( \w + h,\pp  )$,
\begin{subequations}
\label{ehne0}
\beq
\label{eahne0}
\e_{a,h}(\pp) &=& \e(\pp) - h,\\
\e_{b,h}(\pp) &=& \e(\pp) + h,
\eeq
\end{subequations}
where $\e(\pp)$, $\Sig(p)$, $\Pi(p)$, $G(p)$ and $F(p)$  are defined in Eqs.~\eqref{h=0}. 

It is seen from Eq.~\eqref{eahne0} that the spectrum of $a$ particles has a gap which vanishes and the spectrum becomes unstable at $h<h_{c1}(T=0)$, where
\beq
\label{hc10}
h_{c1}(T=0) = \e(\pp_0)
\eeq
and $\pp_{0}$ is the momentum at which $\e(\pp)$ has a minimum. This instability signifies a transition to another phase. One has $\ppp=(\pi,\pi,\pi)$ in DTN because exchange couplings are antiferromagnetic.

One concludes from Eqs.~\eqref{hne0}--\eqref{ehne0} that magnetic field lifts the equivalence between $a$ and $b$ particles. However magnetization 
$M(h,T=0)=\langle S_i^z\rangle$ 
remains zero in the paramagnetic phase as it can be readily seen from Eqs.~\eqref{sz}, \eqref{gahne0} and \eqref{gbhne0}. Thermal fluctuations make finite the magnetization.

\begin{figure}
\includegraphics[scale=1.0]{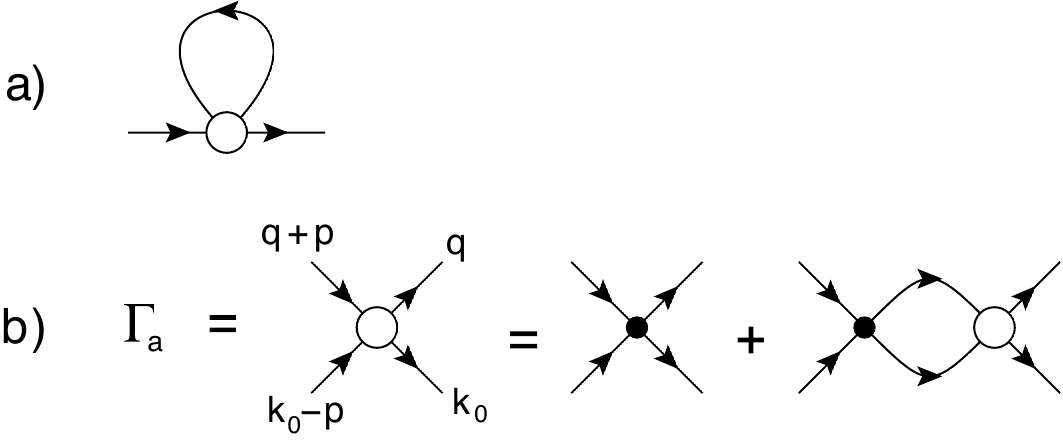}
\caption{
(a) The Hartree-Fock diagram given the first order correction in $J/D$ to the self-energy part.
(b) Diagram equation for the vertex $\Ga_a(\w,\pp,\qq)$ which is involved in the Hartree-Fock diagram. Lines are Green's function \eqref{gfa} of $a$ particles. Black dots are bare vertexes given by Eq.~\eqref{h4a}.
\label{hfmag}}
\end{figure}

\underline{\it $T\ne0$ and $h\sim h_{c1}(0)$.} 
Let us consider the dispersion equation \eqref{dispeq} for $a$ particles. We have found its solution in our previous paper \cite{larged1} at $T=0$ up to the third order in $J/D$. The aim of the present discussion is to find temperature corrections to the spectrum $\delta_T\e_{a,h}(\pp)$ in the first nonvanishing orders in $J/D$ and $T$ considering the temperature to be small enough $T\ll J$. Then, it is convenient to represent the spectrum and self-energy parts using Eqs.~\eqref{hne0} in the following form:
\beq
\e_{a,h}(\pp) &=& \e(\pp) - h + \delta_T \e_{a,h}(\pp),\\
\label{siga}
\Sig_{a,h} (\e_{a,h}(\pp),\pp) &=& \Sig(\e(\pp),\pp) + \left.\frac{\partial \Sig (\w,\pp)}{\partial \w} \right|_{\w = \e(\pp)} \delta_T \e_{a,h}(\pp) + \delta_T \Sig_{a,h} (\e(\pp)-h,\pp), 
\eeq
where $\e(\pp)$ and $\Sig(\w,\pp)$ are the spectrum and the self-energy part at $T=0$, $\delta_T \Sig_{a,h} (\w,\pp)$ is the temperature correction to $\Sig_{a,h} (\w,\pp)$ and expressions for $\Sig_{b,h} (-\e_a(\pp),\pp)$ and $\Pi_h (\e_a(\pp),\pp)$ can be written similar to Eq.~\eqref{siga}. Substituting these equations into Eq.~\eqref{dispeq} and using results of our previous calculation \cite{larged1} of self-energy parts and $\e(\pp)$ up to the third order in $J/D$ we have in the first order in $J/D$
$
\delta_T \e_{a,\pp} = \delta_T \Sig_{a,h} (\e(\pp)-h,\pp)
$
and
$
 h_{c1} (T) = \e(\pp_0) + \delta_T \Sig_a (0,\kkk).
$
The first order correction in $J/D$ to $\delta_T \Sig_a (\e(\pp)-h,\pp)$ is given by the Hartree-Fock diagram shown in Fig.~\ref{hfmag}(a). As a result one obtains
\beq
\delta_T \e_{a,h}(\pp) &=& 4 \Ga_a(\e(\pp)-h,0,\pp)M(h,T),\\
\label{hc1t}
h_{c1}(T) &=& \e(\pp_0) + 4 \Ga_a(0,0,\kkk) M(h,T),
\eeq
where $\Ga_a(\w,\pp,\qq)$ is the vertex and
\beq
\label{mag}
M(h,T) = \frac1N \sum_\kk N(\e(\kk)-h)
\eeq
is equal to the magnetization in the second order in $J/D$ at $T\ll \e(0)+h$. It is explained in our previous paper \cite{larged1} that ladder diagrams give the main contribution to the vertex leading to the Bethe-Salpeter equation for $\Ga_a(\w,\pp,\qq)$ shown in Fig.~\ref{hfmag}(b). To calculate the vertex in the leading order in $J/D$ one can use Green's function in the form $G_a(\w,\pp) = 1/(\w - \e_1(\pp)+h+i\delta)$. When $\w \sim J$, the solution can be tried in the form 
$
\Ga_a(\w, \pp,\qq) = A(\w) + ( J_\pp - J_{\pp+\qq} )/4 + B^z (\w) J^z_{\pp+(\qq-\kkk)/2} + B^{xy} (\w) J^{xy}_{\pp+(\qq-\kkk)/2}.
$
The solution is quite cumbersome and we do not present it here. We point out only that $\Ga_a\sim J$ when $\w\sim J$. We remind also that the value 
\begin{equation}
\label{v0}
v_0=2\Ga_a(0,0,\kkk)
\end{equation}
is an effective two-particle interaction which can be found experimentally at small $T$ as a slope of the plot of $h_{c1}(T)$ vs $M_c(T)=M(h=h_{c1}(T),T)$ (see Eqs.~\eqref{hc1t} and \eqref{mag}).

The specific heat can be obtained from the formula $C(h,T) = d\langle{\cal H}\rangle/dT$ using Eq.~\eqref{ham} with the following result in the first order in $J/D$:
\beq
\label{ct}
C(h,T) = \frac{d}{dT}\left(\frac1N \sum_\kk \Big( \e_1(\kk)-h \Big) N \big( \e_3(\kk)-h \big)\right),
\eeq
where $\e_1(\kk)$ and $\e_3(\kk)$ are spectra at $T=0$ in the first and in the third orders in $J/D$, respectively.

\underline{\it Application to DTN.} Equations for $M(h,T)$, $h_{c1}(T)$ and $C(h,T)$ obtained above are applicable at $h\approx h_{c1}(0)$ only at small enough $T$ so that the thermal corrections to be small. In the case of a quasi-1D system it usually means that the temperature cannot exceed the value of the exchange constant between spin chains ($\sim0.2$~K in DTN). At such $T$ the quasi-1D system behaves like a 3D one and we can expect the proportionality to $T^{3/2}$ of $h_{c1}(T)-h_{c1}(0)$, $M_c(T)=M(h=h_{c1}(T),T)$ and $C(h=h_{c1}(T),T)$ expected for QCP of 3D BEC universality class.

\begin{figure}
\includegraphics[scale=0.5]{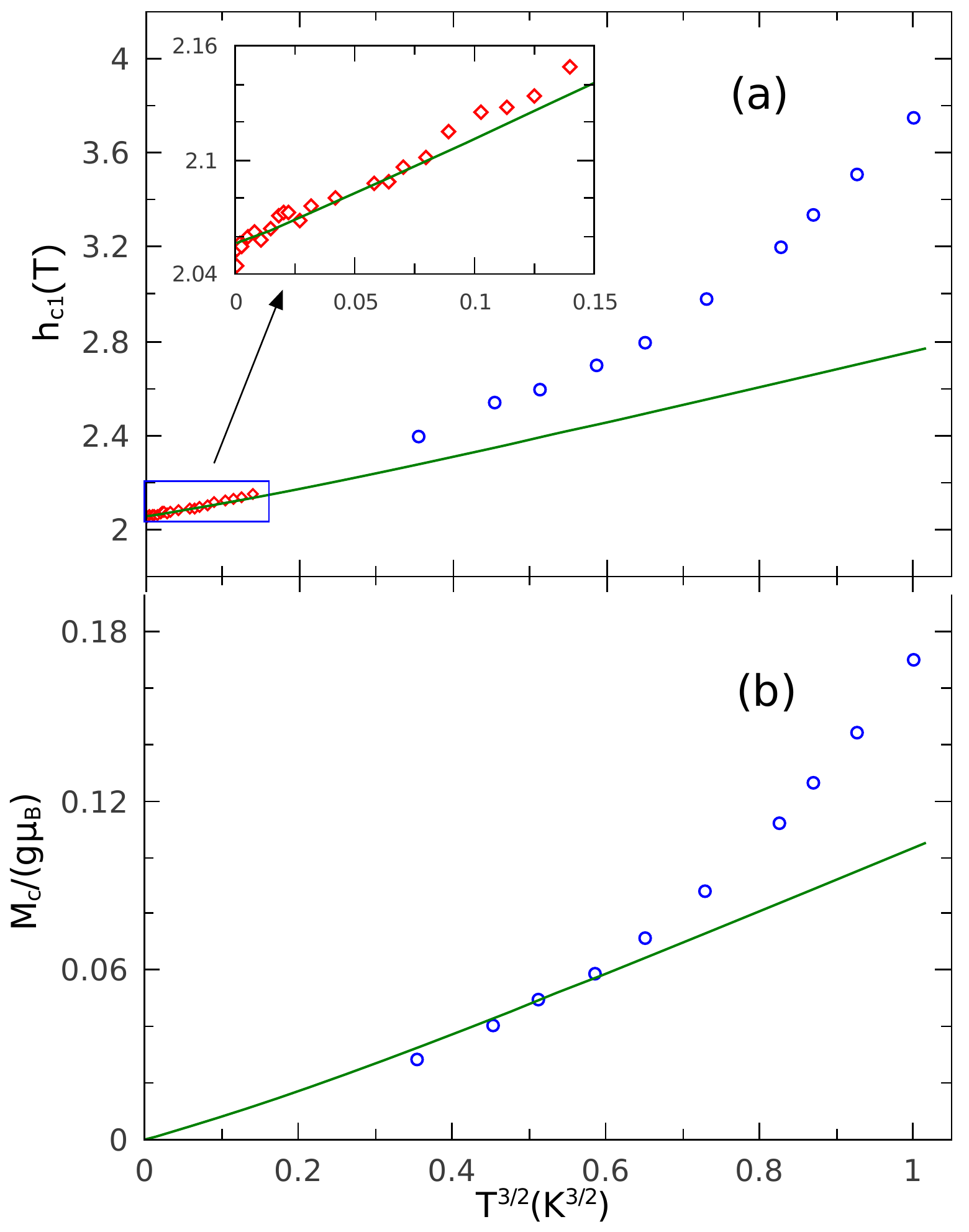}
\caption{(Color online) Plots of (a) $h_{c1}$ and (b) $M_c$ (magnetization at $h=h_{c1}(T)$) vs $T^{3/2}$ in DTN. Circles and diamonds are experimental data from Ref.~\cite{17} and Ref.~\cite{yin}, respectively. Lines are drawn using Eqs.~\eqref{hc1t}, \eqref{mag} and parameters \eqref{ourdtnpar}.
\label{hcmct32}}
\end{figure}

This proportionality was really observed experimentally in DTN. Fig.~\ref{hcmct32} shows the experimental data for $M_c(T)$ and $h_{c1}(T)$ obtained in Ref.~\cite{17} and Ref.~\cite{yin} for $0.5\ {\rm K}<T<1\ {\rm K}$ and $1\ {\rm mK}<T<300\ {\rm mK}$, respectively, together with results of our calculations with Eqs.~\eqref{hc1t}, \eqref{mag} and parameters \eqref{ourdtnpar}. The agreement between the theory and experiment is very good at $T<0.3\ {\rm K}$. Large temperature fluctuations come into play at greater $T$ which are not taken into account in Eqs.~\eqref{hc1t} and \eqref{mag}. As a result the deviation from experimental data is noticeable at $T>0.5\ {\rm K}$ for $h_{c1}(T)$ and at $T>0.7\ {\rm K}$ for $M_c(T)$.

\begin{figure}
\includegraphics[scale=0.5]{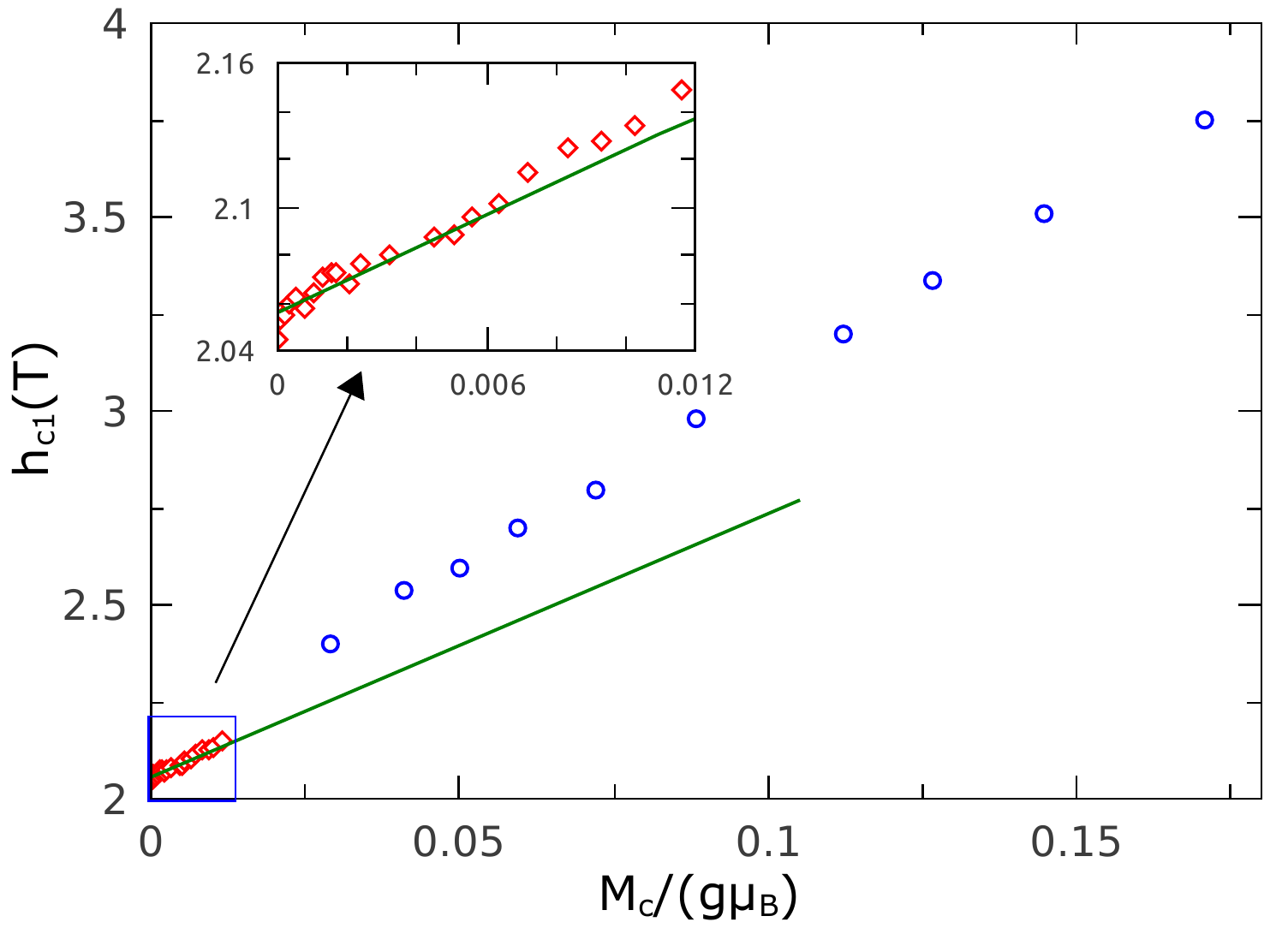}
\caption{(Color online) $h_{c1}(T)$ versus $M_c(T)$ in DTN. Circles are experimental data from Ref.~\cite{17}. Diamonds were putted using experimental data of Ref.~\cite{yin} for $h_{c1}(T)$ and $M_c(T)$ calculated with Eq.~\eqref{mag} and parameters \eqref{ourdtnpar}. Lines are drawn using Eqs.~\eqref{hc1t}, \eqref{mag} and parameters \eqref{ourdtnpar}.
\label{hcmc}}
\end{figure}

Although (and quite expectedly) neither $h_{c1}(T)$ nor $M_c(T)$ do not depend linearly on $T^{3/2}$ at $T>0.7\ {\rm K}$ in DTN (see Fig.~\ref{hcmct32}), it is observed experimentally \cite{17} that $h_{c1}(T)$ is a linear function of $M_c(T)$ at $0.5\ {\rm K}<T<1\ {\rm K}$ as is demonstrated in Fig.~\ref{hcmc}. The effective two-particle interaction is extracted from this plot in Ref.~\cite{17} as it is explained above with the result $v_0\approx0.61$~meV. Most likely, however, that $v_0$ is renormalized by thermal fluctuations in DTN at such large $T$. We plot in Fig.~\ref{hcmc} also $h_{c1}(T)$ vs $M_c(T)$ using the low-temperature experimental data of Ref.~\cite{yin} for $h_{c1}(T)$ and $M_c(T)$ computed from Eq.~\eqref{mag} with parameters \eqref{ourdtnpar}. The effective interaction obtained in this way is equal approximately to 0.44~meV that is 28\% smaller than the value experimentally found in Ref.~\cite{17} at $T>0.5\ {\rm K}$ and that is in excellent agreement with the result of our calculation of $v_0$ by Eq.~\eqref{v0} and parameters \eqref{ourdtnpar}.

\begin{figure}
\includegraphics[scale=0.5]{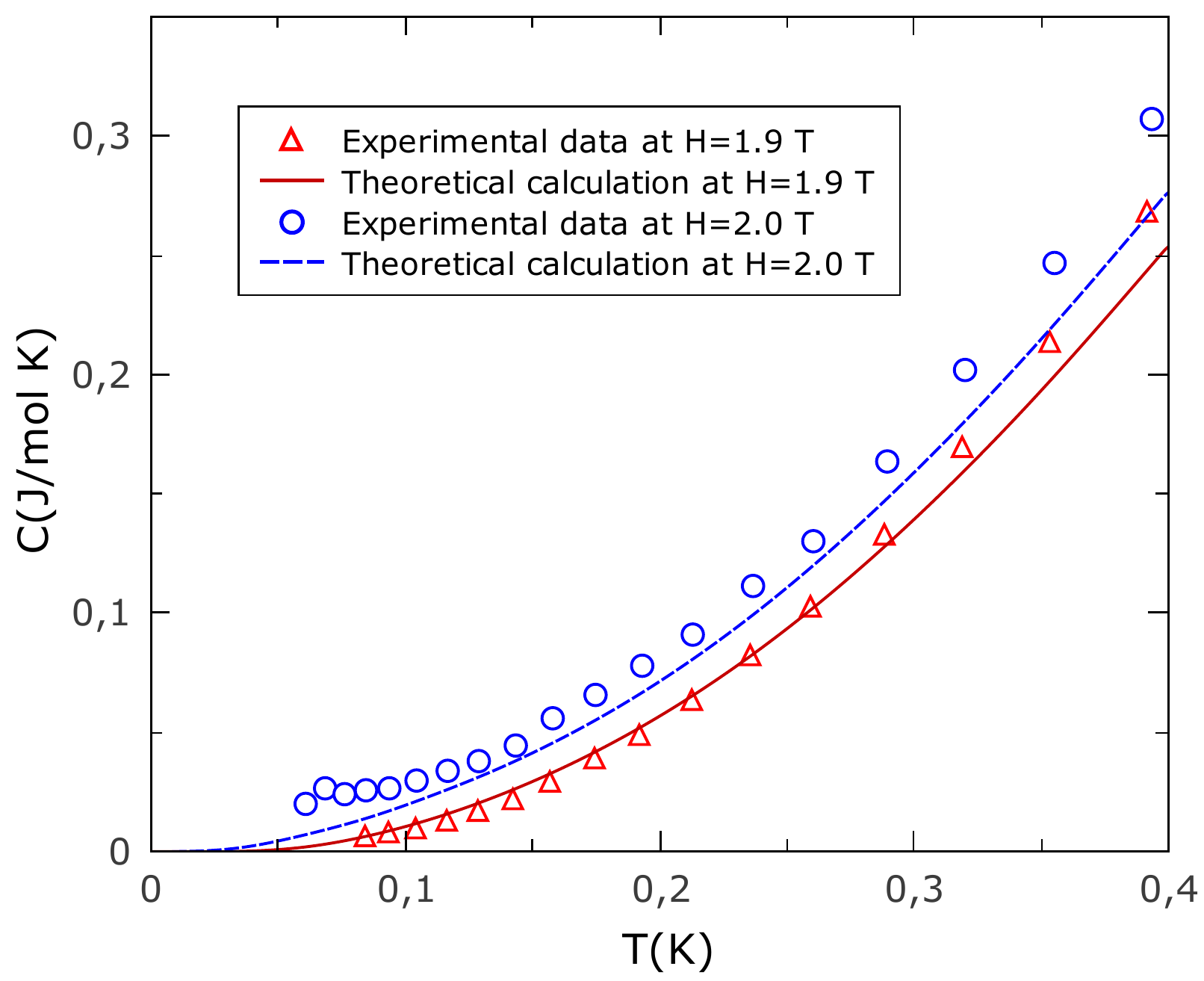}
\caption{(Color online) Specific heat in DTN. Experimental data are taken from Ref.~\cite{chern}. Lines are drawn using Eq.~\eqref{ct} and parameters \eqref{ourdtnpar}.
\label{heat}}
\end{figure}

Experimental data of Ref.~\cite{chern} for $C(h,T)$ at $h\approx h_{c1}(0)$ are shown in Fig.~\ref{heat} together with results of our calculations with Eq.~\eqref{ct}, parameters \eqref{ourdtnpar} and expression for $\e_3(\pp)$ found in Ref.~\cite{larged1}. A reasonable agreement between the theory and experiment is seen at $T<0.3\ {\rm K}$. The specific heat of the model \eqref{ham} with parameters \eqref{ourdtnpar} is proportional to $T^{3/2}$ at $h=h_{c2}(T)$ as well because the spectrum in the fully polarized phase given exactly at $T=0$ by $\varepsilon(\pp)=h-D+J_\pp$ is also quadratic near its minimum. It is demonstrated in Ref.~\cite{chern} that due to the strong renormalization of the spectrum in the paramagnetic phase $C(h=h_{c2}(T),T)$ is about 6 times larger than $C(h=h_{c1}(T),T)$. Simple calculation of the specific heat with the spectrum $\varepsilon(\pp)$ shows that in agreement with the experiment $C(h=h_{c2}(T),T)$ is approximately 5.7 times larger than Eq.~\eqref{ct} at $h=h_{c1}(T)$ at small $T$.

This work was supported by RF President (grant MK-329.2010.2) and RFBR grant 09-02-00229.

\bibliography{LargeD} 

\end{document}